# Frequency tunable topological edge states of two-dimensional honeycomb lattice photonic crystals

**Yuchen Peng, Bei Yan, Jianlan Xie, Exian Liu, Hang Li, Rui Ge, Feng Gao and Jianjun Liu**∗

*Key Laboratory for Micro/Nano Optoelectronic Devices of Ministry of Education & Hunan Provincial Key Laboratory of Low-Dimensional Structural Physics and Devices, School of Physics and Electronics, Hunan University, Changsha 410082, China*

**∗ Corresponding Email: jianjun.liu@hnu.edu.cn**

**Abstract**-- In this paper, the photonic quantum spin Hall effect (PQSHE) is realized in dielectric two-dimensional (2D) honeycomb lattice photonic crystal (PC) by stretching and shrinking the honeycomb unit cell. Combining two honeycomb lattice PCs with a common photonic band gap (PBG) but different band topologies can generate a topologically protected edge state at the combined junction. The topological edge states and their unidirectional transmission as the scatterers with triangular, pentagonal, and heptagonal shapes are researched. Meanwhile, the unidirectional transmission in an inverted Ω-shaped waveguide with large bending angle is realized, and verifies the characteristics of the topological protection by adding different kind of defects. Moreover, the frequency varies significantly when changing the scatterers shape, which shows that the PC with various scatterers shape can tune the frequency range of the topological edge state significantly. In other words, it can adjust the frequency of unidirectional transmission and increase the adjustability of the topological edge state.

# I. Introduction

In recent years, topological photonics has become a hot scientific research field [1-7], because topological photonic devices and systems can generate many novel and excellent properties, such as unidirectional transmission and defect immunity without loss. In 2008, Haldane and Raghu firstly proposed a method for realizing the quantum Hall effect by applying strong magnetic field to break time-reversal (TR) symmetry in magneto-optic PC [8,9]. Subsequently, some researchers used this method to design a mass of PC unidirectional waveguides and related devices [10-21].

However, the magnetic materials in the magneto-optical PC have large volume, with the remarkable absorption loss, which are mainly applied to the microwave frequency band, so researchers put their eyes on the dielectric materials. In 2015, Hu *et al.* [22] realized the PQSHE attributed to spin-orbit coupling protection that preserves TR symmetry by using a 2D honeycomb lattice dielectric PC composed of $C_{6v}$ symmetric honeycomb unit cell. A new method to generate topological state is proposed by stretching and shrinking the honeycomb lattice to exchange bands order. Moreover, Yang *et al.* [23] experimentally realized the PQSHE in the 2D honeycomb lattice PC in the millimeter wave band. Later, there are a series of research on the PQSHE using the honeycomb lattice model, such as, using air hole scatterers [24-26], triangular scatterers [24,26], realizing asymmetric radiation of quantum dots by PQSHE [26], covering the surface of the dielectric cylinder with a layer of graphene [27], the waveguide confined between two layers of graphene slabs [28], liquid crystal material injected into the gap of the dielectric cylinders [29]. Therefore, the topological edge states of 2D honeycomb lattice PCs based on dielectric materials have been widely researched in depth. In addition, compared to other structures with $C_{6v}$ symmetry (such as, the core-shell [30] and Stampfli-type [31] PC with triangular lattice), 2D honeycomb lattice PC structure parameters are easier to adjust. Furthermore, double Dirac cone in band structure is not accidental degeneracy point [32] and the appearance of its topological edge states is more regular.

It is well known that the scatterers shape and size can affect the filling ratio of the PC, which in turn affects the effective relative permittivity and bands structure of the PC [33]. Moreover, the generation and characteristics of the topological edge states are closely related to the band structure [22-30]. Therefore, the scatterers shape and size can ultimately tune the needful frequency range of topological edge states for the unidirectional transmission of the PC waveguide. It has been reported that the change of scatterers size (the size of air hole, the inner and outer diameter size of core-shell scatterers) can generate band inversion [25,30] and realized topological edge states. Therefore, the influence of scatterers shape on topological edge states and unidirectional transmission effect are still an interesting and relevant problem to be addressed.

In this paper, the band structure and topological edge states of a 2D honeycomb lattice PC are investigated by the finite element method when the scatterers shapes are triangle, pentagon and heptagon respectively, which provides a method for finding common PBG and realizing unidirectional transmission within different frequencies. In other words, it can offer more clues to find and tune topological edge states.

Moreover, frequency tunable topological edge states offer the prospect for developing the dynamical control of waveguide with robust transmission.

## II. Model and Theory

The structure of 2D honeycomb lattice PC is shown in Fig. 1.

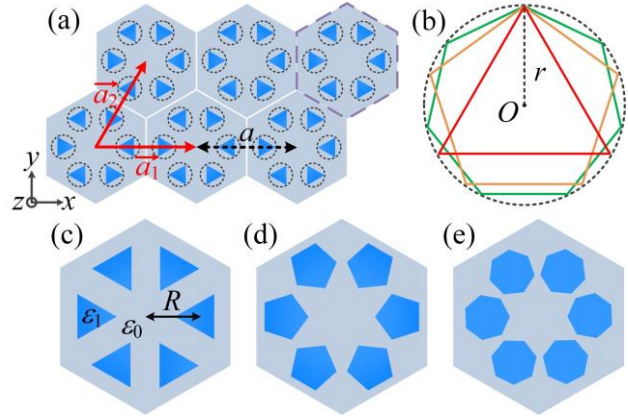

FIG. 1. (a) Honeycomb unit cell (purple dotted line) arranged in a triangular lattice to form a 2D honeycomb lattice PC; (b) Various shapes scatterers and their circumscribed circle; honeycomb unit cell with various shapes scatterers: (c) Triangle; (d) Pentagon; (e) Heptagon.

As shown in Fig. 1(a), the lattice vectors of the 2D honeycomb lattice PC are $\vec{a}_1 = (a, 0)$ and $\vec{a}_2 = (a/2, \sqrt{3}a/2)$, where the lattice constant is set as $a$=1μm. There are six regular polygon scatterers in each honeycomb unit cell, and the distance between the unit cell center and the center of scatterers is $R$. The lattice and unit cell both have $C_{6v}$ symmetry. In this paper, the scatterers material is Si ($\varepsilon_r$=11.9) and the background material is air ($\varepsilon_r$=1). In order to avoid the overlap of the regular polygon scatterers in the 2D space, set the circumscribed circle as shown in Fig. 1(b). If the circumscribed circles do not overlap, it can be ensured that the regular polygon scatterers will not overlap. The radius of the circumscribed circle is $r$=0.12μm. The honeycomb unit cells with triangular, pentagonal, and heptagonal scatterers are shown in Figs. 1(c), 1(d), and 1(e), respectively.

It is known that a triangular lattice structure satisfying the $C_{6v}$ symmetry can realize a double Dirac cone in bands [34], that is, satisfying the quadruple degeneracy condition. The Dirac cones at the K (K′) point are folded to form a double Dirac cone in the Brillouin zone (BZ) center (Γ points) when taking the honeycomb unit cell [24]. By continuously changing R, the double Dirac cone is opened, and the p-d bands inversion will occur, which can further lead to topological phase transitions between nontrivial and trivial states.

According to the $k \cdot p$ perturbation theory, effective Hamiltonian $H_{\text{eff}}$ for the photonic bands near the Γ point can be derived. The Bloch functions must satisfy the following eigenvalue equation [22]

$$\frac{1}{\varepsilon(\vec{r})} \nabla \times \nabla \times \vec{H}_{n\vec{k}}(\vec{r}) = \left(\frac{\omega_{n\vec{k}}}{c}\right)^2 \vec{H}_{n\vec{k}}(\vec{r}) \qquad (1)$$

In Eq. (1), c is the speed of light, is the position-dependent relative permittivity, $\varepsilon(\vec{r})$ is the magnetic field vector, while the indices n and $\vec{H}_{n\vec{k}}$ are the band number and the wave vector respectively. An effective Hamiltonian in the form of a 4×4 matrix representation can be obtained when the Bloch functions is expanded at finite k to the BZ center and restricted on the four p and d states. It is known that $p_x$, $p_y$, $d_{x^2+y^2}$, $d_{xy}$ modes can form four kinds of pseudospin states [22]

$$\begin{aligned} |p_+\rangle = |p_x + ip_y\rangle, \quad |d_+\rangle = |d_{x^2-y^2} + id_{xy}\rangle \\ |p_-\rangle = |p_x - ip_y\rangle, \quad |d_-\rangle = |d_{x^2-y^2} - id_{xy}\rangle \end{aligned} \qquad (2)$$

Then, the effective Hamiltonian $H_{\text{eff}}$ in the representation $(|p_+\rangle, |d_+\rangle, |p_-\rangle, |d_-\rangle)^T$ can be derived after a series of calculations [30]

$$H_{\text{eff}} = \begin{pmatrix} \frac{\omega_p^2}{c^2} & Ak_+ & 0 & 0 \\ A^*k_- & \frac{\omega_d^2}{c^2} & 0 & 0 \\ 0 & 0 & \frac{\omega_p^2}{c^2} & A^*k_- \\ 0 & 0 & Ak_+ & \frac{\omega_d^2}{c^2} \end{pmatrix} \quad (3)$$

where $k_\pm = k_x \pm i k_y$, $A = \langle p_+|H|d_+\rangle = \langle p_-|H|d_-\rangle$, which is the $k \cdot p$ coupling coefficient between different states, and $\omega_p$ and $\omega_d$ are the eigen-frequency of the $p$ band and $d$ band at the $\Gamma$ point respectively. The TR symmetry requires $A$ must be a pure imaginary number, and the case of double Dirac cone at the $\Gamma$ point is $\omega_p = \omega_d = \omega_0$, when the group velocity is $|A|c^2/2\omega_0$. Obviously, only the same spin polarization direction (i.e., $p_+$ and $d_+$, $p_-$ and $d_-$) will be coupled. It can be found that the effective Hamiltonian $H_{\text{eff}}$ has a similar form to the Bernevig-Hughes-Zhang (BHZ) model in the CdTe/HgTe/CdTe quantum well system which describes the quantum spin Hall effect. The total Berry phase near the Dirac cone is zero, so the system is called to preserve TR symmetry. Based on Eq. (3), the spin Chern numbers [35,36] can be derived

$$C_\pm = \pm \frac{1}{2}\left[\text{sgn}(M) + \text{sgn}(B)\right] \quad (4)$$

It can be known from Eq. (4) that when $MB>0$, $C_\pm = \pm 1$, the system is in the topological nontrivial state; when $MB<0$, $C_\pm = 0$, the system is in the topological trivial state. Where $B$ is the diagonal element of the second term of $k \cdot p$ perturbation in effective Hamiltonian $H_{\text{eff}}$ and is typically negative. $M=(E_d-E_p)/2$, where $E_d$ and $E_p$ are the energy of $d$ and $p$ bands respectively, indicating that the positive and negative of $M$ can be changed as long as the $p$ bands and $d$ bands are exchanged, which is the reason why bands inversion can generate topological phase transition.

## III. Results and Discussion

The band structure of the 2D honeycomb lattice PC with three various shapes scatterers and their $E_z$ field distributionof the $p$ and $d$ bands are calculated, as shown in Fig. 2.

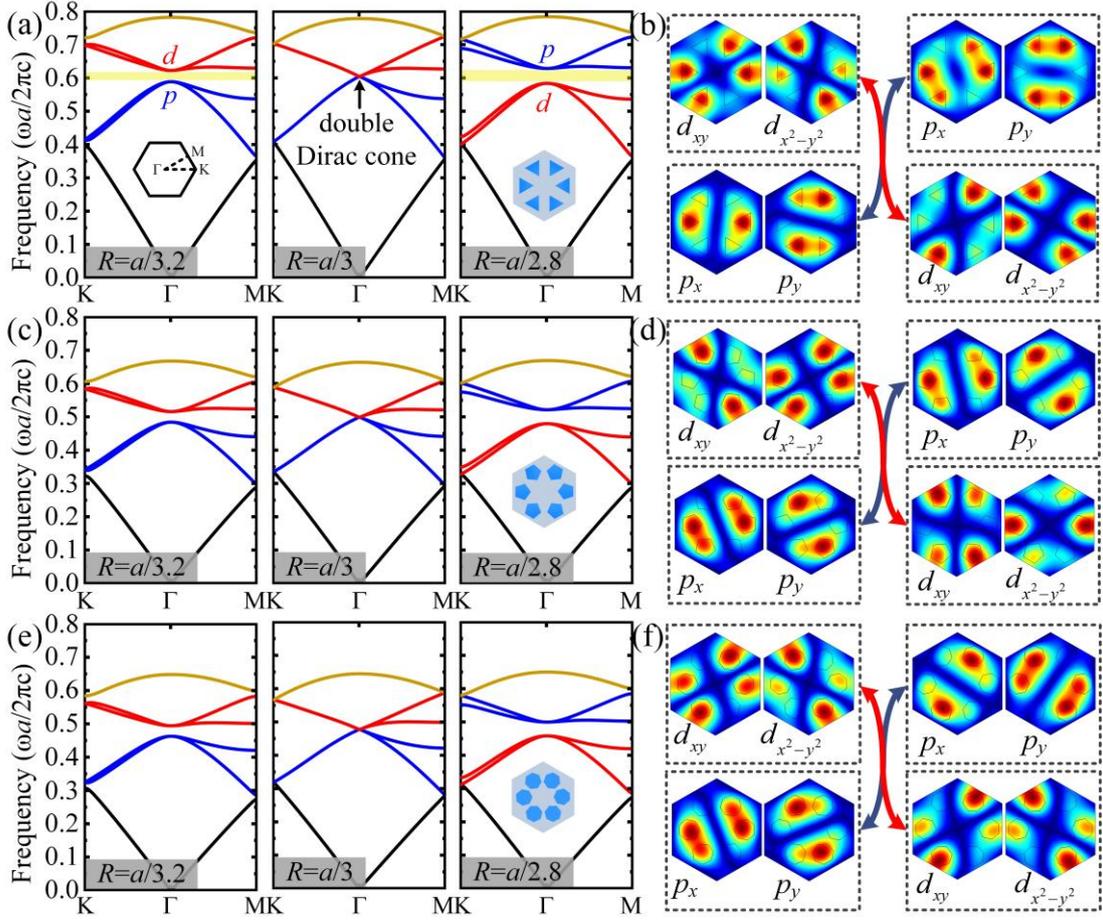

FIG. 2. The band structure of the 2D honeycomb lattice PC with various shapes scatterers and their $E_z$ field distribution. The blue and pink bands represent $p_\pm$ and $d_\pm$ bands respectively: Triangular scatterers: (a) Band structure; (b) $E_z$ field distribution; Pentagonal scatterers: (c) Band structure; (d) $E_z$ field distribution; Heptagonal scatterers: (e) Band structure; (f) $E_z$ field distribution.

The prerequisite for PQSHE is that the PCs of the two topological states share a common PBG. The structure with $C_{6v}$ symmetry can open the Dirac cone by changing the size of $R$. At this time, the $p$-$d$ bands inversion generates a topological phase transition. It can be seen from the band structures of Figs. 2(a), 2(c), and 2(e) that, the 2D honeycomb lattice PC realizes a double Dirac cone when $R=a/3$. The $p_\pm$ bands have a pair of dipole modes, and the $d_\pm$ bands have a pair of quadrupole modes near the Γ point. In each of the $E_z$ field distributions of Figs. 2(b), 2(d), and 2(f), the left side pattern corresponds to the field distribution for the case of $R=a/3.2$, and the right side pattern corresponds to the field distribution for the case of $R=a/2.8$, the upper pattern represents the field distribution corresponding to the 4$^{th}$ and 5$^{th}$ bands, and the lower pattern represents the field distribution corresponding to the 2$^{nd}$ and 3$^{rd}$ bands. From this, it can be seen that when $R$ is changed from $a/3.2$ to $a/2.8$, the phenomenon of $p$-$d$ bands inversion occurs in the 2D honeycomb lattice PC. Comparing Figs. 2(a), 2(c) and 2(e), the change of the scatterers shape can tune the frequency of the Dirac cone and the PBG. Moreover, a law can be obtained that, as the edge number of scatterers increases, the PBG shifted downward gradually. Therefore, changing the scatterers shape provides a way to find a common PBG, and realize unidirectional transmission within different frequency ranges by changing the scatterers shape.

The reason why the PBG is gradually shifted downward is that, changing the scatterers shape changes the filling ratio $f_c$ and the effective relative permittivity $\varepsilon_{\text{reff}}$. The filling ratio is the ratio of scatterers area to the unit cell area. Therefore, according to mathematical derivation, the filling ratio of honeycomb lattice PC with $N$-edge scatterers is

$$f_c = 2\sqrt{3} \cdot N \cdot \left(\frac{r}{a}\right)^2 \cdot \sin\frac{2\pi}{N} \quad (N=3,4,5,...) \tag{5}$$

It can be seen from Eq. (5) that the filling ratio is related to the edge number of the regular polygon and the radius of the circumscribed circle of the scatterers. When the radius of the circumscribed circle $r$ is determined, the filling ratio $f_c$ increases as the edge number *N* increases. At the same time, changing the filling ratio will change the effective relative permittivity $\varepsilon_{\text{reff}}$, which can be obtained by the Maxwell–Garnett relation [37]

$$\varepsilon_{\text{reff}} = \varepsilon_{rb} + 3f_c \varepsilon_{rb} \frac{(\varepsilon_{ra} - \varepsilon_{rb})}{\varepsilon_{ra} + 2\varepsilon_{rb} - f_c(\varepsilon_{ra} - \varepsilon_{rb})} \tag{6}$$

where $\varepsilon_{ra}$, $\varepsilon_{rb}$ are the relative permittivity of the scatterers and the background material respectively. It can be seen from the Eq. (6) that as $f_c$ increases, $\varepsilon_{\text{reff}}$ gradually increases. Liu *et al.* [38] showed that the PBG frequency gradually shifted downward with the increase of the effective permittivity, which can justify the observed phenomenon in this paper. It can be seen from Fig. 2 that the change trend of the PBG is consistent with the change of Dirac cones. Therefore, in order to make the law become more explicit, the relationship between the frequency of the Dirac cone and the number of edges is plotted, as shown in Fig. 3.

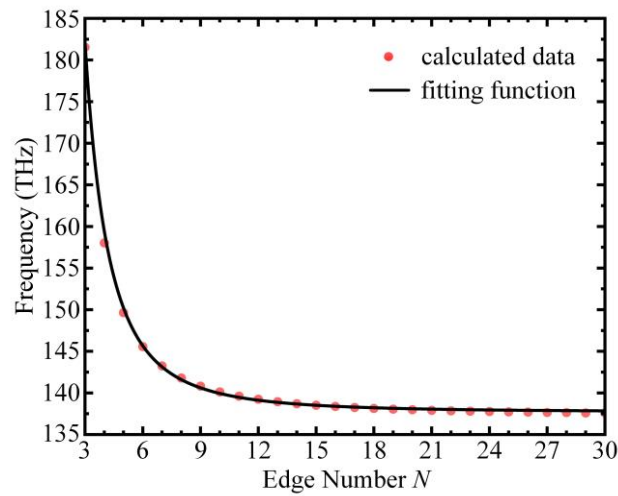

FIG. 3. The relationship between the frequency of the Dirac cone and the number of scatterers edge.

It can be seen from the plot that, frequency varies significantly when *N* is no more than 10. The function between the frequency of the Dirac cone $f$ and the number of edges $N$ can be deduced through numerical fitting

$$f = 676.5N^{-2.479} + 137.7 \tag{7}$$

whose RMSE (Root Mean Squared Error) is 0.2692. When $N$ approaches infinity, $f$ is 137.7±0.2692Thz, which indicates the predicted frequency of the Dirac point when the scatterers are circular, and the actual frequency is 137.28Thz. The difference is very minor, demonstrating that this function is reasonable.

As is apparent from Figs. 2 and 3, changing the scatterers shape can change the frequency of each band, but cannot change the band structure. That is, changing the scatterers shape only tunes the frequency of the topological edge states, but does not change the generation and characteristics of the topological edge states. In order to avoid redundancy, the topological edge states and their unidirectional transmission effects in 2D honeycomb lattice PCs are calculated and analyzed by taking the pentagonal scatterers as an example.

The basic evidence of the band topology is the topologically protected edge state at the interfaces can be generated when combining the topological trivial state and the topological nontrivial state PCs [25], which can be calculated by the supercell method along the $k_x$ direction. The projected band and its electric field distribution are shown in Fig. 4.

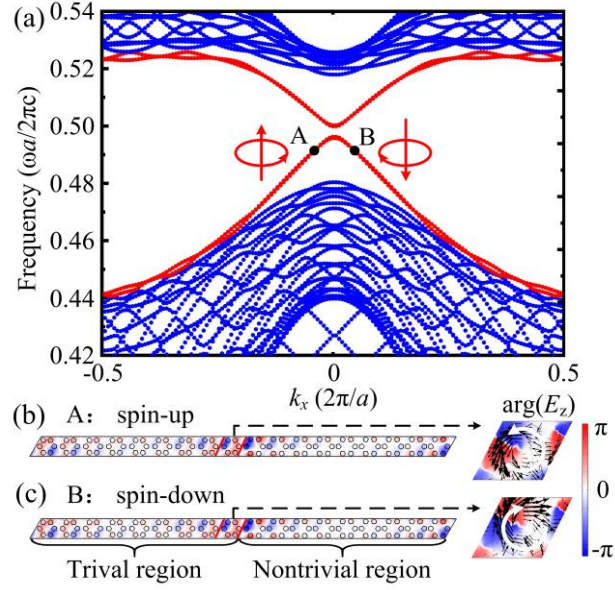

FIG. 4. Topological edge states and supercells: (a) The projected band of PCs with pentagonal scatterers of two different topological states; (b) $E_z$ fields in real-space, phase distribution and energy flow density vector at point A; (c) $E_z$ fields in real space, phase distribution and energy flow density vector at point B.

The projected band as shown in Fig. 4(a) is calculated by the structure of nine $R=a/3.2$ topological trivial unit cells and nine $R=a/2.8$ topological nontrivial unit cells. The red line located in the PBG represents the topologically protected edge state with spin-locked, and the normalized frequency of the topological edge states is $\omega_5 \in [0.478(2\pi c/a), 0.523(2\pi c/a)]$, which is consistent with the PBG frequency in Fig. 2(c). Figs. 4(b) and 4(c) show the energy flow density vector (time-averaged Poynting vector) (indicated by the black arrows) and the phase near the interfaces between the PCs with topological trivial and the topological nontrivial states (between the red lines). The phase distribution $\arg(E_z)$ shows that the edge states locate at the junction and decay in the bulk. Through the direction of the energy density vector and the phase distribution $\arg(E_z)$, it can be determined that the point A (B) represents the pseudospin up state (pseudospin down state), whose frequency is in the vicinity of $\omega=0.495(2\pi c/a)$. In the same way, the projected band, the energy flow density vector and phase distribution of PCs with triangular and heptagonal scatterers are obtained, but their frequencies are different. In the case of a triangular and heptagonal scatterers, the normalized frequencies of the topological edge states of the 2D honeycomb lattice PCs are $\omega_3 \in [0.584(2\pi c/a), 0.631(2\pi c/a)]$, $\omega_7 \in [0.455(2\pi c/a), 0.501(2\pi c/a)]$ respectively. It can be seen that the frequency of topological edge state is changed significantly.

In addition, a new phenomenon is found that, the band with topological edge state will become a flat band with near-zero group velocities when several scatterers are replaced, which can be used to design a slow light waveguide, as shown in Fig. 5.

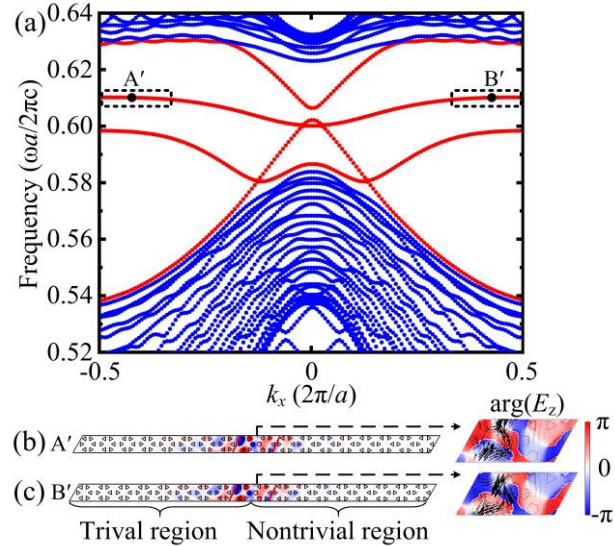

FIG. 5 (a) Topological edge states of PCs with triangular scatterers when four triangular scatterers are replaced by four pentagonal scatterers near the boundary; (b) $E_z$ fields in real space, phase distribution and energy flow density vector at point A′; (c) $E_z$ fields, phase distribution and energy flow density vector in real space at point B′.

It can be seen from Fig. 5 that, when several scatterers are replaced near the boundary between two PCs with different topological states, the originally overlapping edge states will split up into four separate edge states. Furthermore, the band with edge state becomes a flat band with near-zero group velocities in the vicinity of $0.612(2\pi c/a)$, as shown in the rectangle boxes in Fig. 5, meaning that slow light waveguide can be implemented by changing the shape of the scatterers, except of changing the arrangement of the scatterers [39]. This method is an excellent starting point for further realizing slow light effect in PC waveguides.

In order to verify the topological edge states of 2D honeycomb lattice PC, an inverted Ω-shaped waveguide with large bending angle is designed based on PC with pentagonal scatterers. Moreover, its unidirectional transmission effect is shown in Fig. 6.

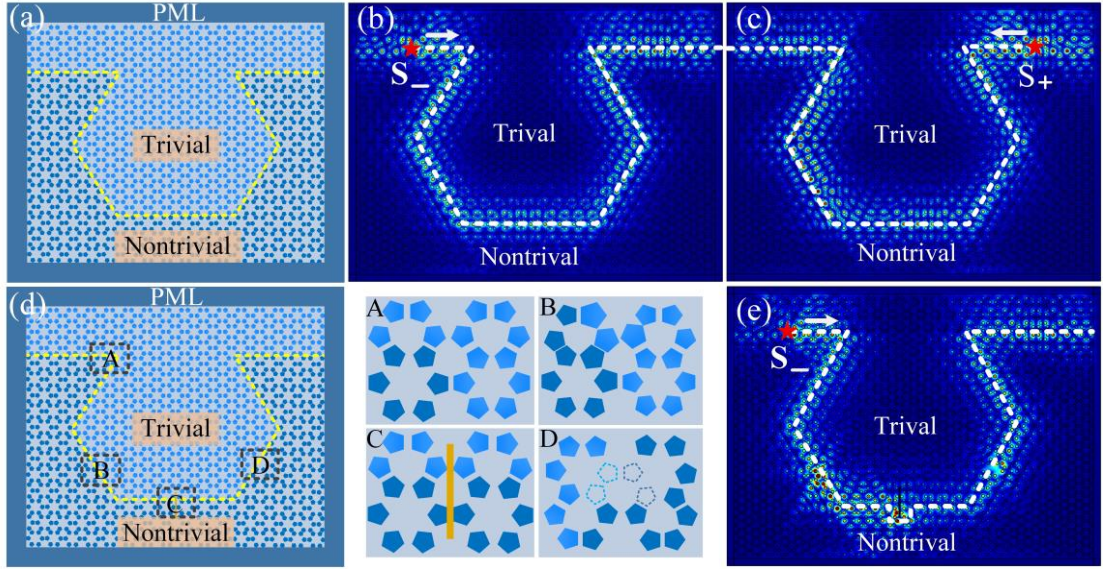

FIG. 6. PC waveguide with pentagonal scatterers transmits against multiple large bending angles in one direction, the light source is marked as a red star, and the $S_+$ ($S_-$) represents the pseudospin up (pseudospin down): (a) Schematic of inverted Ω-shaped waveguide surrounded by perfect matching layer (PML); (b) Unidirectional transmission to the right; (c) Unidirectional transmission to the left; (d) Schematic of inverted Ω-shaped waveguide after adding different kind of defects, which contains A: large bend, B: enlarged scatterers, C: Cu baffle and D: cavity; (e) Unidirectional transmission to the right after adding different kind of defects.

In Fig. 6(a), the upper (lower) component of the waveguide is composed of a topological trivial (nontrivial) PC where $R=a/3.2$ ($R=a/2.8$). The perfect matching layer around the waveguide absorbs electromagnetic waves, the light source $S_\pm$ is $H_0 e^{i\omega t}(\hat{x} \mp i\hat{y})$, where $H_0$ is the magnetic field of arbitrary amplitude, and $\omega$ is the eigen frequency. This light source can generate electromagnetic waves with a counterclockwise/clockwise circularly polarized surface magnetic field. An out-of-plane electric field with a positive/negative angular momentum wave function is generated. As shown in Figs. 6(b) and 6(c), the PC waveguide can realize both the rightward and the leftward unidirectional transmission, whose frequency is 155.194THz, which can smoothly pass through multiple large bending angles, proving its robust unidirectional transmission effect and topological protection. The inverted Ω-shaped waveguide after adding some different defects is shown in Fig. 6(d). It can be seen from Fig. 6(e) that the defects in the waveguide have almost no influence on the unidirectional transmission effect, that is, the 2D honeycomb lattice PC can overcome obstacles and disorders to realize unidirectional transmission with strong robustness. In the same way, the unidirectional transmission effect and robustness of the 2D honeycomb lattice PC with the triangular, heptagonal and pentagonal scatterers are consistent, but the frequencies are different. In the case of 2D honeycomb lattice PC waveguides with triangular and heptagonal scatterers, the unidirectional transmission frequencies can be 176.843THz and 140.802THz, respectively. Therefore, frequency tunable topological edge state can realize dynamical control of waveguide with different unidirectional transmission frequency.

## IV. Conclusion

In this paper, the topological edge states of 2D honeycomb lattice PCs with triangular, pentagonal and heptagonal scatterers are analyzed. It can be found that as the edge number of scatterers increases, the frequency of the topological edge states gradually decreases. Moreover, the frequency varies significantly when the edge number of scatterers is no more than 10. 2D honeycomb lattice PCs with polygon scatterers can realize unidirectional transmission of inverted Ω-shaped waveguide with large bending angle. Meanwhile, a phenomenon is found that only replacing several scatterers of PCs can realize slow light waveguide. In conclusion, changing the scatterers shape can realize unidirectional transmission with different frequency and overcome large bending angle with strong robustness, which provides potential applications for integrated photonic devices with tunable frequency and optical communication systems with robust transmission.

## Acknowledgements

This work was supported by the National Natural Science Foundation of China (Grant No. 61405058), the Natural Science Foundation of Hunan Province (Grant No. 2017JJ2048), and the Fundamental Research Funds for the Central Universities (Grant No. 531118040112). The authors acknowledge Professor Jianqiang Liu for software sponsorship.